\documentclass{pasj00}
\draft

\begin{document}
\SetRunningHead{Author(s) in page-head}{Running Head}

\title{Digital Spectro-Correlator System for the Atacama Compact Array
of the Atacama Large Millimeter/submillimeter Array}
\author{
Takeshi \textsc{Kamazaki}\altaffilmark{1}
Sachiko K. \textsc{Okumura}\altaffilmark{1}
Yoshihiro \textsc{Chikada}\altaffilmark{1}
\\
Takeshi \textsc{Okuda}\altaffilmark{1,2}
Yasutaka \textsc{Kurono}\altaffilmark{1}
Satoru \textsc{Iguchi}\altaffilmark{1}
\\
Shunji \textsc{Mitsuishi}\altaffilmark{3}
Yuji \textsc{Murakami}\altaffilmark{3}
\\and\\
Naomitsu \textsc{Nishimuta}\altaffilmark{3}
Haruo \textsc{Mita}\altaffilmark{3}
Ryo \textsc{Sano}\altaffilmark{3}
}

\altaffiltext{1}{
National Astronomical Observatory of Japan, Mitaka, Tokyo
181-8588, Japan
}
\email{kamazaki.takeshi@nao.ac.jp}
\altaffiltext{2}{
Department of Astrophysics, Nagoya University, Nagoya, Aichi
464-8602, Japan
}
\altaffiltext{3}{
Fujitsu Advanced Engineering, Shinjuku, Tokyo
163-1018, Japan
}


\maketitle

\begin{abstract}
 We have developed an FX-architecture digital spectro-correlator for the
 Atacama Compact Array (ACA) of the Atacama Large
 Millimeter/submillimeter Array. The correlator is able to
 simultaneously process four pairs of  dual polarization signals with
 the bandwidth of 2 GHz, which are received by up to sixteen
 antennas. It can calculate auto- and cross-correlation spectra
 including cross-polarization in all combinations of all the antennas,
 and output correlation spectra with flexible spectral configuration
 such as  multiple frequency ranges and multiple frequency
 resolutions. Its spectral dynamic range is estimated to be
 higher than $10^{4}$ relative to $T_{sys}$ from processing results of
 thermal noise for eight hours with a typical correlator
 configuration. The sensitivity loss is also confirmed to be 0.9 \% with
 the same configuration. In this paper, we report the detailed design of
 the correlator and the verification results of the developed hardware.
\end{abstract}

\section{Introduction}

\begin{figure}
 \begin{center}
  \FigureFile(85mm,85mm){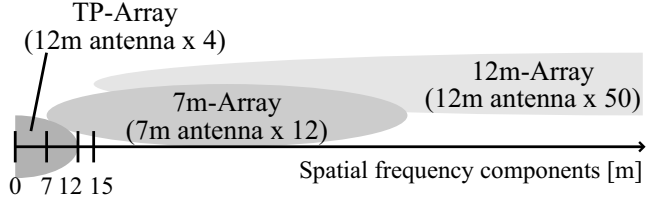}
 \end{center}
 \caption{ALMA consists of two different types of arrays: 12m-Array to
 obtain spatial frequency components higher than 15 m; and
 Atacama Compact Array (ACA), composed of TP-Array and 7-m Array, to
 cover zero- and short-baseline components.}
 \label{fig:alma}
\end{figure}

The Atacama Large Millimeter/submillimeter Array (ALMA), an
international astronomy facility, is a partnership of Europe, East Asia,
and North America in cooperation with the Republic of Chile. ALMA is a
general-purpose radio telescope with 66 antennas or more at a high
altitude site (about 5100 m) in the Atacama Desert of northern Chile
\citep{Beasley2006}.
ALMA intends to provide precise images of all kinds of
astronomical objects in the wavelength range from 0.3 mm to 3 mm and
achieve angular resolution of \timeform{0.01''}.
ALMA consists of the 12m-Array composed of fifty or more 12-m antennas
and the Atacama Compact Array (ACA) composed of twelve 7-m antennas
(7m-Array) and four 12-m antennas (Total power (TP)-Array) (see
Figure \ref{fig:alma}). The purpose of the 12m-Array is to obtain
spatial frequency components between 15 m (= 12 m $\times$ close packing
ratio (= 1.25)) and the maximum baseline length toward target objects by
interferometric method. The ACA mainly intends to obtain short
spatial components including those of zero-baseline length in
interferometric and single-dish observations with the 7m-Array and the
TP-Array, respectively. The obtained components by the three
arrays are combined to construct high fidelity images of the target
objects. The images lack little for spatial information between
zero-baseline and the maximum baseline lengths \citep{Iguchi2009}.


The correlator is a key element to realize high fidelity imaging of
various astronomical targets. In aperture synthesis observations,
astronomical signals received at the antennas are correlated to produce
astronomical fringes. Total power is derived from auto-correlation of
the signals received in single-dish observations.
Since the correlator plays an important role in this processing,
we have developed a new FX method correlator for the ACA
(hereafter, referred to as ``ACA Correlator''), which calculates Fourier
transform (denoted as ``F'') and then performs cross-multiplication
(denoted as ``X''). For the 12m-Array, NRAO and European groups
developed a different type of correlator called ``64-Antenna
Correlator'' (\cite{Escoffier2007}).

At the start of our development, the number of Fast Fourier Transform
(FFT) points was in the order of 10000 at most. However, ALMA requires
the highest frequency resolution of 5 kHz or less for 2 GHz IF-band,
which indicates that about one million FFT points are necessary. There
was no FX correlator with such a huge number of FFT points at the
time. Hence, as a first step, we developed a prototype FX correlator in
order to verify its design and implementation \citep{Okumura2002}, as
well as basic performance prior to the development of the ACA
Correlator.

In this paper, we describe specifications and requirements in Section
\ref{sec:specs_reqs}, design in Section \ref{sec:design}, hardware
implementation in Section \ref{sec:hw_implementation}, and function and
performance verification in Section \ref{sec:verification}.

\section{ALMA specifications and requirements for the ACA Correlator}
\label{sec:specs_reqs}

The main purpose of the ACA is to provide short-baseline and total-power
data of various astronomical objects with high precision. These data are
combined with longer-baseline data obtained by the 12m-Array for high
fidelity imaging of astronomical objects \citep{Iguchi2009}. As the ACA
is sometimes used as a single array for observations, it is also
required to have reasonable imaging capabilities \citep{Iguchi2009}.
Thus, the ACA Correlator should have (1) the capability to
process receiver signals from up to 16 ALMA ACA antennas, (2) the high
processing performance sufficient to calculate a large quantity of
correlation spectra among all the ACA antennas, (3) the high precision
processing enough to achieve high spectral dynamic range and (4) a low
sensitivity loss. In addition to them, there are requirements for
correlator functions, sub-array mode and full-array mode, data reception
of digitized receiver signals, delay tracking in a long
baseline, sideband processing, spectral configuration, temporal
integration, compatibility with the 64-Antenna Correlator and operations
at the ALMA array operation site. The details of ALMA
specifications and requirements for the ACA Correlator are summarized in
Table \ref{tbl:acafx_spec}.


\begin{table}
 \caption{ALMA specifications and requirements for the ACA Correlator}
 \label{tbl:acafx_specs_reqs}
 \begin{center}
  \begin{tabular}{ll}
   \hline \hline
   Input & \\
   \hline
   Number of sub-arrays & 2 (TP-Array/7m-Array) \\
   \hline
   Number of antennas   & 16 \\
   \hline
   \raisebox{1.5ex}{Number of inputs per antenna}
   & \shortstack[l]{\\
   4 pairs \\
   $\times$ [2 GHz bandwidth IF band $\times$ 2 polarizations]} \\
   \hline
   \raisebox{1.5ex}{Data}
   & \shortstack[l]{\\
   3-bit 4 Gsps optical signal \\
   (Common format between the ACA and the 12m-Array)} \\
   \hline \hline
   Output per IF band pair & \\
   \hline
   Number of auto-correlation   & 16 antennas \\
   \hline
   Number of cross-correlation  & 120 baselines \\
   \hline
   Number of polarization & 4 polarizations \\
   \hline \hline
   \multicolumn{2}{l}{Performance} \\
   \hline
   \raisebox{2.5ex}{Spectral dynamic range}
   & \shortstack[l]{\\
   10000:1 for weak spectral lines near stronger ones \\
   1000:1 for weak spectral lines on strong continuum \\ emission} \\
   \hline
   Sensitivity loss & $\le$ 8.3 \% \\
   \hline
   Baseline length & $\le$ 15 km \\
   \hline \hline
   \multicolumn{2}{l}{Compatibility with the 64-Antenna Correlator} \\
   \hline
   Sideband processing    & sideband separation, sideband rejection \\
   \hline
   \raisebox{2.5ex}{Spectral configuration}
   & \shortstack[l]{\\
   up to 32 frequency ranges (31.25 MHz - 2 GHz) \\
   highest frequency resolution of 5 kHz or less in \\
   31.25 MHz bandwidth } \\
   \hline
   \raisebox{1.5ex}{Temporal integration}
   & \shortstack[l]{\\
   1 ms for auto-correlation only output \\
   16 ms for auto- and cross-correlation output} \\
   \hline \hline
   Operation at the high site & about 5100 m above sea level \\
   \hline
  \end{tabular}
 \end{center}
\end{table}

\paragraph*{Processing of auto- and cross-correlation spectra:}
The ACA Correlator is required to have the ability to process auto- and
cross-correlation (including cross-polarization) in all combinations of
all the ACA antennas.

\begin{figure}
 \begin{center}
  \FigureFile(74mm,85mm){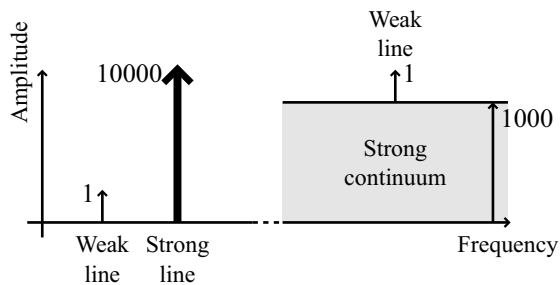}
 \end{center}
 \caption{ALMA has two spectral dynamic requirements: 10000:1 for the
 measurement of weak spectral lines near stronger ones and 1000:1 for
 weak lines in the presence of strong continuum emission.}
 \label{fig:alma_spdyn}
\end{figure}

\paragraph*{Spectral dynamic range:}
In the auto- and cross-correlation processing, high precision processing
is required to achieve high spectral dynamic ranges of ALMA. Its
specifications are 10000:1 for the measurement of weak spectral lines
near stronger ones and 1000:1 for weak lines in the presence of strong
continuum emission (see Figure \ref{fig:alma_spdyn}).


\paragraph*{Sensitivity loss:}
Low sensitivity loss smaller than 12 \% is required for the ALMA digital
system including digitizer and correlator. Since ALMA adopts a 3-bit
digitizer whose sensitivity loss is 3.7 \%, the allowable sensitivity
loss is less than 8.3 \% within the ACA Correlator.

\paragraph*{Sub-array mode and full-array mode:}
TP-Array shall be independently operated as a single-dish telescope to
obtain total power data, while 7m-Array shall be operated as an
interferometer to obtain short-baseline data. Two arrays shall also
operate together as a single array (Full-Array) for high
accuracy calibration. Hence, the ACA Correlator is required to have two
operational modes: the sub-array mode to perform simultaneous
data processing of  TP-Array and 7m-Array and the full-array mode to
operate two arrays as a single array.


\begin{figure}
 \begin{center}
  \FigureFile(78mm,85mm){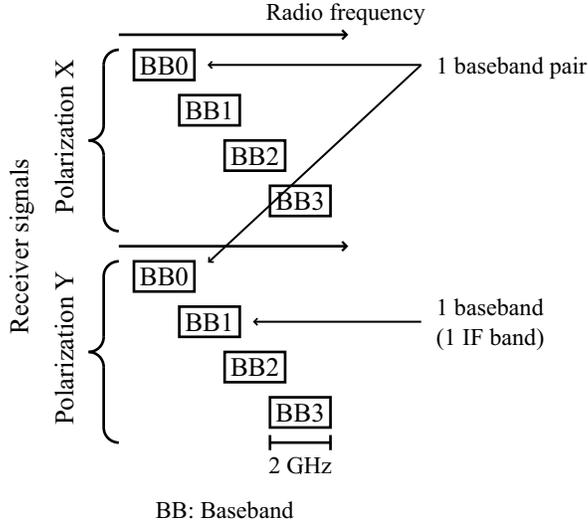}
 \end{center}
 \caption{ALMA can select four IF bands paired in different
 polarizations from receiver signals, which pairs are  referred to
 baseband pairs.}
 \label{fig:alma_baseband}
\end{figure}

\paragraph*{Reception of digitized receiver signals:}
ALMA has dual polarization receivers, where four IF bands of 2 GHz
bandwidth are available for every polarization in each observational
band (see Figure \ref{fig:alma_baseband}). The single IF band
is referred to as a baseband, and two basebands paired in different
polarizations are called a baseband pair. The data of four baseband
pairs, whose total bandwidth is 16 GHz, are digitized by 3-bit 4 Gsps
digitizers and sent to the ACA and 64-Antenna Correlators through
optical fibers with common digital format between the ACA and the
12m-Array. The ACA Correlator should be capable of receiving
the formatted data from 16 ACA antennas.

\paragraph*{Baseline length:}
ALMA spreads its antennas at the array operation site, and the longest
distance between the antennas reaches about 15 km
(\cite{Beasley2006}). Since the arrival time of a signal wavefront from
a target differs from antenna to antenna, the delay of the wavefront 
should be compensated before the signals are correlated. Thus,
the delay compensation should be frequent enough to keep the
sensitivity loss to 8.3 \% or less in the ACA Correlator.

\paragraph*{Sideband processing:}
ALMA adopts a two-single sideband (2SB) receiver system between Band 3
(84 - 116 GHz) and Band 8 (385 - 500 GHz), because the 2SB system has
advantages in spectral-line observations, polarization observations and
calibration observations from the viewpoint of signal-to-noise ratio
(\cite{Iguchi2005}). For Band 9 (602 - 720 GHz) and Band 10 (787 - 950
GHz), the DSB system is adopted because of its
high technical difficulty in developing the 2SB system.
Thus, the ACA Correlator requires sideband rejection
capability including sideband separation by 90-degree phase switching or
LO offset (see Section \ref{subsec:lo_offset}). The sideband processing
in the 2SB system is also effective to improve the sideband rejection
ratio, which is only 7 dB in ALMA requirements.

\paragraph*{Spectral configuration:}
In spectroscopic observations with ALMA, it is necessary to perform
spectroscopy of an appropriate IF band frequency range with an
appropriate frequency resolution, because required frequency range and
resolution vary depending on astronomical objects to be
observed. Hence, ALMA requires flexible spectral configurations
up to 32 frequency ranges with different frequency resolutions in every
baseband pair. These multiple frequency ranges are called sub-bands in
ALMA. The widest bandwidth of a sub-band is 2 GHz and the narrowest is
31.25 MHz. The highest frequency resolution is 5 kHz or less in a
sub-band of 31.25 MHz bandwidth.  It is also required that these
sub-bands can be overlapped by another.

\paragraph*{Temporal integration:}
According to the ALMA requirements, the ACA Correlator shall be
capable of performing 1 ms and 16 ms temporal integration
in the mode to output auto-correlation only and both of
auto- and cross-correlation, respectively. Further temporal integration
is performed by post-processing computers.

\paragraph*{Compatibility with the 64-Antenna Correlator:}
Functional and data compatibility between the ACA Correlator and the
64-Antenna Correlator is very important, because the frontend, backend
and LO subsystems are all common to the ACA and the 12m-Array, and their
output data are combined as previously mentioned. Thus, the ACA
Correlator is required to provide a mode compatible with the 64-Antenna
Correlator for sideband processing, spectral configuration, and temporal
integration.

\paragraph*{Operations at the high site:}
The correlator hardware shall be designed by taking maintainability and
cooling function into account, because it is installed at the high site
at about 5100 m above sea level.

\section{Design}
\label{sec:design}

\begin{figure}
 \begin{center}
  \FigureFile(44mm,85mm){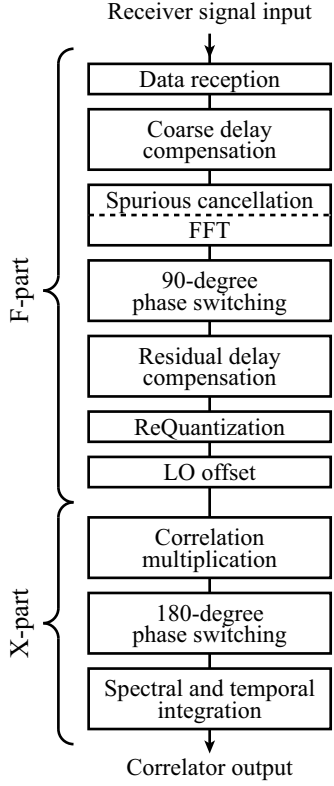}
 \end{center}
 \caption{Processing flow of the ACA Correlator. The flow
 consists of an F-part and X-part. The F-part  mainly processes data
 reception, FFT and re-quantization, while the X-part mainly processes
 correlation multiplication and spectral and temporal integration.}
 \label{fig:acafx_flow}
\end{figure}

According to the ALMA specifications and requirements for the
ACA Correlator described in the previous section, we decided
on a correlation method and designed a processing flow and correlator
functionality in detail. The processing flow of the ACA
Correlator is shown in Figure \ref{fig:acafx_flow}.

\subsection{Correlation method}

For the ACA Correlator, we selected a conventional FX method
\citep{Chikada1987} from several methods, i.e., FX, XF and their
enhanced methods.
Normal FFT in the fixed-point arithmetic has been studied in analysis
and simulation, and its characteristics are well known. Thus, the normal
FFT is deemed to be suitable to realize low sensitivity of 8.3 \% or
less and a high spectral dynamic range of 10000:1 required in
ALMA. In addition, the FX correlators have the advantage of providing
more flexible spectral configuration than other methods. Since the
conventional FX correlator always performs spectroscopy of input signals
using Fourier transform in the highest frequency resolution, it is easy
to output requested multiple frequency ranges with necessary frequency
resolutions by choosing the highest frequency resolution spectra and
summing them up. Having such flexibility is important for general
purpose telescopes such as ALMA. Another advantage is frequency
response of sinc squared function as a result of the calculation
sequence of ``F'' and ``X'' \citep{Kamazaki2008a}. The response has
relatively steep edge and low sidelobe level among the methods. This
feature is suitable for precise imaging, which is one of the main goals
of ALMA.

Regarding the FX method, two enhanced methods have been proposed: a
polyphase FX correlator by \citet{Bunton2003} and an FFX correlator by
\citet{Iguchi2008}.
The polyphase FX correlator uses a polyphase filter bank instead
of a normal FFT and the filter bank can suppress spectral
leakage better than normal FFT. The polyphase FX correlator can also
provide multiple  frequency resolutions by cascading the polyphase
filter banks. 
On the other hand, the FFX correlator performs a 2-stage
Fourier transform. The 1st stage Fourier transform provides spectra of
full bandwidth, and the 2nd stage Fourier transform calculates spectra
with higher frequency resolution from several parts of the full
bandwidth spectra. The 1st stage Fourier transform functions as a
digital filter bank. The developed FFX correlator has been used in the
Atacama Submillimeter Telescope Experiment (ASTE).
These two methods can also provide multiple frequency ranges with
necessary frequency resolutions required in ALMA. However, their
available ranges and resolutions depend on the cascading ways
and performance of basic processing units (polyphase filter bank for the
polyphase FX method and FFT for the FFX method). This suggests that the
polyphase FX and FFX methods are more complex in correlator hardware
composition and less flexible in spectral configuration compared to the
conventional FX correlator. Thus, we decided to adopt the conventional
FX method, although it requires more computational performance than the
two enhanced FX methods.

On the other hand, the 64-Antenna Correlator adopts an enhanced
XF method. The conventional XF method performs correlation (denoted as
``X'') and then Fourier transform (denoted as ``F''). Hence, frequency
resolution depends on the number of available correlation lags and
bandwidth, which means that the highest frequency resolution is usually
achieved with the narrowest bandwidth only. To compensate for the
inflexibility in the spectral configurations, the 64-Antenna Correlator
is equipped with a digital tunable filter bank. Such a type of
correlators are called an FXF correlator after its filter bank (denoted
as ``F''). In this method, signals are filtered by each bandpass filter
in the filter bank, and separately processed using the normal XF
method. Thus, it can provide more flexible spectral
configurations than the conventional XF correlator within the
limits of the digital filter bank and the number of available
correlation lags.


\subsection{Coarse delay compensation}
\label{subsec:coarse_delay}

Correlation between different antennas shall be performed using the same
signal wavefront from the target astronomical object. However, the
arrival time of the same wavefront differs from antenna to antenna
depending on their locations. Cancelling out such time difference is
called ``delay compensation''.
The ACA Correlator performs delay compensation in two methods: coarse
delay compensation and residual delay compensation. The coarse delay
compensation is a method to shift digitized data in the time domain
according to the time difference. Since the data is shifted by a unit of
a sample, the minimum step is 250 ps (= 1 sample / 4 GHz). Remaining
delay error smaller than 1 sample is compensated in the residual delay
compensation after FFT.

If the delay error is serious, the phase slope heavily inclines in the
frequency domain. As a result of this, sensitivity loss is caused by
phase rotation within one frequency channel. To prevent the loss, it is
necessary to perform the compensation frequently. The maximum
sensitivity loss is roughly estimated by $1 - \sin(2\pi \Delta\nu
\Delta\tau)/(2\pi \Delta\nu \Delta\tau)$. $\Delta\nu$ is the maximum
frequency offset from the phase rotation center, which is 4 GHz in the
ALMA. $\Delta\tau$ is the maximum delay change, which is 3.6 ns s$^{-1}$
$\times$ [interval of delay compensation (sec)] at the baseline length
of 15 km. Thus, the ACA Correlator performs both of the two delay
compensations at an interval of 1 ms in order to keep the sensitivity
loss to less than 1 \% at the maximum baseline length of 15 km.

\subsection{Detection and cancellation of low-frequency and
  high-frequency spurious signals}
\label{subsec:spurious_signal_cancellation}

To meet the scientific requirements for frequency resolution, the ACA
Correlator adopts a FFT segment length of 1,048,576(=$2^{20}$;
hereafter described as 1M) points. Using a long segment length allows
observations with very high frequency resolution, but at the same time
increases susceptibility to Electro-Magnetic Interference (EMI) and DC
offset arising in the vicinity of Analog-to-Digital converters
(digitizer), because the spurious signals are amplified by 1024 times
compared to the system noises through FFT calculations. To avoid harmful
overflows caused by them, high frequency spurious (HFS) cancellation and
low frequency spurious (LFS) cancellation are implemented by the ACA
Correlator.

The digitizers are placed in an environment where high-frequency EMI is
present, such as reference clocks, their harmonics and their cross
modulated waves. Since the EMI is synchronous to a digitizer clock, they
form a ``fixed'' pattern waveform, whose length is a multiple of a
digitizer clock period and a reciprocal of the lowest frequency of such
EMI. These spurious signals may cause overflows in the fixed-point
arithmetic FFT and degrade output data connected to the overflow points
through the FFT butterfly network. In order to avoid the
overflows in FFT operations, the ``fixed'' pattern waveform of the HFS
signal should be measured and subtracted from the input data before the
FFT operations.
To subtract the HFS signal, the ``fixed'' pattern waveform is measured
before the FFT segmentation as follows: Input i-th ($i=0,1,...$)
time-series data are accumulated into the j-th bin, where $j$ is the
remainder of $i$ divided by the fixed pattern length, until a
sufficiently accurate pattern is obtained. The required accuracy is
determined by the required spectral dynamic range, which is
1/1000 or less in voltage (corresponding to -60 dB relative to
$T_{SYS}$). Thus, the sufficient accuracy, 1/1000 root-mean-square of
the input data, will be achieved by 2$^{20}$ times accumulation every
bin. After this accumulation, the bins values are normalized by the
number of accumulation ($2^{20}$) and subtracted by the average across
the whole pattern in order to nullify the DC component in the pattern
(see Equation \ref{eq:hfsc}).
The pattern length is selected to be 960 (= $2^{6} \times 3 \times 5$)
in order to subtract sin waves synchronous to the digitizer clock, whose
frequencies are any integer weighted sum of the combination of the
following frequencies;
4000/2,   4000/3,   4000/4,   4000/5,   4000/6,   4000/8,   4000/10,
4000/12,  4000/15,  4000/16,  4000/20,  4000/24,  4000/30,  4000/32,
4000/40,  4000/48,  4000/60,  4000/64,  4000/80,  4000/96,  4000/120,
4000/160, 4000/192, 4000/240, 4000/320, 4000/480, 4000/960 (MHz).

\begin{eqnarray}
 H'_{j} & = & \frac{1}{2^{20}}\sum_{n=0}^{(2^{20}-1)} d_{j+960n} \nonumber \\
 H_{j} & = & H'_{j} - \frac{\sum_{k=0}^{(960-1)} H'_{k}}{960} \label{eq:hfsc} \\
 j & : & 0,1,2, ... , (960-1) \nonumber \\
 d_{i} & : & \mbox{$i(=j+960n)$-th sample of time-series data} \nonumber \\
 H_{j} & : & \mbox{HFS signal pattern} \nonumber
\end{eqnarray}

Another EMI may arise due to electric leaks from power supply lines and
magnetic fluxes leaked from cooling fan motors. Since the frequency of
such EMI is lower than the highest frequency resolution (3.815 kHz) of
the ACA Correlator, this type of EMI affects the DC channel of 1M-point
FFT. The DC offset of the digitizer also contributes to the DC
channel. Leakage of the digitizer clock itself is aliased and may also
appear in the DC channel. This may cause overflow in the FFT
operation. Thus, the DC offset of the LFS signal should be also
subtracted from the input data before the FFT operations.
The LFS offset is measured by averaging input time-series data within an
FFT segment after the FFT segmentation (see Equation \ref{eq:lfsc}). The
LFS detection provides sufficient signal-to-noise ratio, which is 1/1000
of root-mean-square of the input data and the same as that of HFS
detection, with an FFT segment. For this reason, the LFS signal is
detected in real time every FFT segment, and its cancellation is
performed for the detected FFT segment.
Note that 180-degree phase switching also suppress the DC offset, but
does not prevent the 1M-point FFT from overflow. This is because the
180-degree phase switching subtracts DC offset after temporal
integration over its switching complete cycle, which is performed after
the correlation multiplication following the 1M-point FFT.

\begin{eqnarray}
 L & = & \frac{1}{2^{20}}\sum_{i=0}^{(2^{20}-1)} d_{i} \label{eq:lfsc} \\
 d_{i} & : & \mbox{$i$-th sample of time-series data} \nonumber \\
 L & : & \mbox{LFS signal} \nonumber
\end{eqnarray}

\subsection{FFT}

The FFT is required to be performed at 1,048,576(=$2^{20}$) or
more points in order to achieve the highest frequency resolution
of 5 kHz in the 2 GHz bandwidth. The 1M spectral points FFT
provides the highest frequency resolution of 2 GHz / 512K frequency
channels (= 1M-point / 2; K$\equiv 2^{10}$) = 3.815 kHz, which is higher
than 5 kHz required by ALMA and is equivalent to the highest frequency
resolution of the 64-Antenna Correlator. Besides, there is no
disadvantage in terms of compatibility between the ACA and the
12m-Array. Thus, we decided to adopt 1M spectral points FFT.

The bit length of the FFT calculations is driven by the
required spectral dynamic range and sensitivity loss. The
spectral dynamic range of 10000:1 requires that the calculation
bias within the ACA Correlator be less than -60 dB relative to
$T_{SYS}$, assuming strong astronomical signals of 1/100 $T_{SYS}$ at
the highest frequency resolution. This suggests 10-bit or more accuracy
is needed in the FFT calculations, while longer bit length is preferable
to decrease calculation noise caused by number rounding. For these
reasons, 16-bit fixed-point FFT calculation is adopted.
Number rounding also affects the spectral dynamic range and sensitivity,
but this is necessary because finite bit length is used in the digital
processing. Truncation or clip of unnecessary bits in the number
rounding adds some calculation bias and noise to original values and,
consequently has an adverse impact on the spectral dynamic range and
sensitivity. The bias is avoidable by using ``convergent rounding''. In
this method, a value whose fraction is not 0.5 is rounded to its nearest
integer, and a value with a fraction of 0.5 is rounded to even integer
nearest to the value. The ACA Correlator adopts this method
where it is available. However, since the noise is inevitable, the
correlator adopts as long bit-length as possible.

In the 1M-point FFT operation, 4-Gsps time series data of finite length
is used as input data, and each data segment is called FFT
segment. Since  4-Gsps data is used to obtain correlation data of 1 or
16 ms integration corresponding to 4 $\times$ $10^{6}$ or 64 $\times$
$10^{6}$ samples, 1M-point FFT is necessary to be initiated every 250
$\mu s$ or 10$^{6}$ samples. This indicates that 1M-point FFT lacks
48576 (= 2$^{20}$ - 10$^{6}$) samples. To make up the difference, two
approaches were discussed: to pad zeros for the data shortage;
and to have adjacent FFT segments overlapped. If zeros are
padded for the data shortage, it means to apply rectangular window
function, whose length is 10$^{6}$ points to FFT segment. This introduces
convolution of FFT results of the rectangular function to the original
FX frequency response profile, which is sinc$^{2}$ function. In order to
avoid such degradation, adjacent FFT segments are overlapped by about 5
\% of a segment length.

\subsection{90-degree phase switching}
\label{subsec:90d_phase_sw}

The ALMA system adopts 180-degree phase switching for bias and
spurious cancellation and 90-degree phase switching for sideband
rejection and sideband separation. Since the LO system, where phase is
modulated, is common to the ACA and the 12m-Array, it is required to
apply the same phase switching specifications to them.
The switching base time and switching pattern length of the 180-degree
phase switching are 125 $\mu s$ and 128, respectively. Since the
180-degree and 90-degree phase switchings are nested, the switching base
time of the 90-degree phase switching is equal to the switching
completion cycle of the 180-degree phase switching, which is 16 ms (=
125 $\mu s$ $\times$ 128 patterns). Phase switching patterns are common
to the 90-degree and 180-degree phase switching, and hence, switching
completion cycle of the 90-degree phase switching is 2048 ms (= 16 ms
$\times$ 128 patterns).

128 Walsh functions are candidate patterns for these phase switchings
(\cite{Emerson2006}). However, the ACA cannot freely choose patterns
from them, because switching base time needs to be a multiple of FFT
operation interval (250 $\mu s$) of the ACA Correlator in order to avoid
sensitivity loss by a phase change within a FFT segment. In addition,
the pattern length of 128 is not necessary, because the number of the
ACA antennas is only 16. Thus, we choose available Walsh
functions for the ACA with the following two conditions. The first is
that they are two or four consecutive patterns with an effective
switching base time
of 250 $\mu s$ (= 125 $\mu s$ $\times$ 2)
or 500 $\mu s$(= 125 $\mu s$ $\times$ 4).
The second is that the switching pattern length is 
32 (=16 x 2 consecutive patterns),
64 (=32 x 2 consecutive patterns or 16 x 4 consecutive patterns)
or 128 (=32 x 4 consecutive patterns)
with an effective switching complete cycle of 16 or 32. Detailed
results are summarized in \citet{Kamazaki2008b}.

\subsection{Residual delay compensation}
\label{subsec:residual_delay}

The residual delay compensation is used to compensate delay errors
smaller than 1 sample. In the residual delay compensation, phase slope
of voltage spectra in the frequency domain, which is caused by residual
delay error, is compensated by increasing or decreasing spectral phase
in proportion to both of the delay error and the frequency offset from
tracking frequency.

\subsection{Re-quantization}
\label{subsec:re-quantization}

The 1M-point FFT outputs 512K 16-bit complex values (voltage spectra)
every baseband of 16 antennas. All of the values are needed to be
transferred to the X-part to calculate all correlation spectra over
2-GHz bandwidth by correlation multiplication. The total required
data rate of a baseband pair is 16-bit $\times$ 2 (real- and
imaginary-part of the complex values)
$\times$ 512K frequency channels
$\times$ 2 polarizations
$\times$ 16 antennas / 250$\mu s$
(effective time length of 1 FFT segment)
$\sim$ 2150 Tbps.
In order to reduce the data rate from the F-part to the X-part
and hardware size of the X-part, the complex values are re-quantized to
4-bit real and 4-bit imaginary numbers at the output of the F-part.
The word length of the re-quantization is determined by a trade-off
between possible data deterioration and impact of hardware size.
Receiver signals in ALMA are already deteriorated by the 3-bit
digitizer, which adds noise of about 3.7 \% of the signal power
\citep{Iguchi2002}. A 3, 4 or 5-bit re-quantizer adds another 3.7 \%,
1.1 \% or 0.34 \%, respectively, which were estimated by re-quantization
simulation using a Monte Carlo method. If there is no other
noise in the correlator, the selection of 3-bit, 4-bit or 5-bit
corresponds to a sensitivity loss of 7.3 \%, 4.8 \% and 4.0 \%
or increase of observation time of 16 \%, 10 \% and 9 \% compared to the
case without the re-quantization, respectively.
The hardware cost of a correlation multiplier is approximately
proportional to the square of the word length at its input. In
terms of total size of the ACA Correlator, increase of several
percentages is expected by adoption of 5-bit re-quantization instead of
4-bit re-quantization. Since the 3-bit choice would
reduce the sensitivity too much and a 5-bit solution
would not be affordable, we decided to choose 4-bit solution.

\subsection{LO offset}
\label{subsec:lo_offset}

\begin{figure}
 \begin{center}
  \FigureFile(121mm,85mm){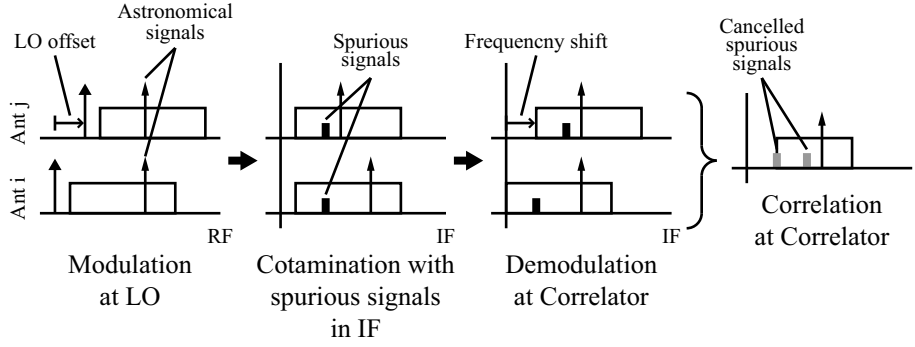}
 \end{center}
 \caption{The LO offset is also used to cancel unwanted signals and
 sidebands. The unwanted signals and sideband, whose frequencies are
 differently shifted among antennas, are cancelled by correlation
 operations.}
 \label{fig:lo_offset}
\end{figure}

Another method to suppress spurious signals and unwanted sidebands,
is to adopt the LO offset in addition to the 180-degree and 90-degree phase
switching.
An LO offset adds slightly shifted frequency offset to
the LO frequency of
the antennas (modulation) and subtracts the frequency offset before
correlation (demodulation) as shown in Figure \ref{fig:lo_offset}. Since
the frequency of an unwanted sideband and common spurious signal
mixed between the
modulation and the demodulation is differently shifted at each antenna
by the demodulation, the unwanted sideband and the spurious signals can
be cancelled by correlation operation.

The ACA Correlator realizes the demodulation of the LO offset by shifting
frequency channels after 1M-point FFT and before correlation
multiplication. The minimum step of the frequency channel shift is 30.5
kHz (= 3.815 kHz $\times$ 8 frequency channels), which is the same as
that of the 64-Antenna Correlator. The maximum value is about 39.1 MHz.


\subsection{Correlation multiplication}

Correlation multiplication of every pair of voltage spectra is
calculated by complex multiplication, which multiplies one complex value
by the complex conjugate of the other complex value. In the
case of different antennas or different polarizations of the same
antenna, cross-correlation spectrum (cross-power spectrum) is produced,
while in the case of the same antenna and the same polarization,
auto-correlation spectrum (auto-power spectrum) is obtained by the
correlation multiplication of identical voltage signals.

\subsection{180-degree phase switching}
\label{subsec:180d_phase_sw}

180-degree phase switching is also implemented for bias and spurious
cancellation (see Section \ref{subsec:90d_phase_sw} for details). Note
that 180-degree phase switching is not effective in suppressing DC
offsets and spurious signals in higher frequency resolution of the
64-Antenna Correlator, because maximum lags (time offsets) between two
antennas (e.g. $\pm 62.5 \mu s$ in the frequency resolution of 8 kHz)
becomes comparable with the switching base time (125 $\mu s$), and the
timing offset caused by the lags can not be ignored between their two
Walsh functions. This is an inherent problem of the XF-method
correlators, however it can be avoided by setting the switching base
time to a multiple of FFT segment length with the FX-method correlators.

\subsection{Spectral and temporal integration}

\begin{figure}
 \begin{center}
  \FigureFile(85mm,85mm){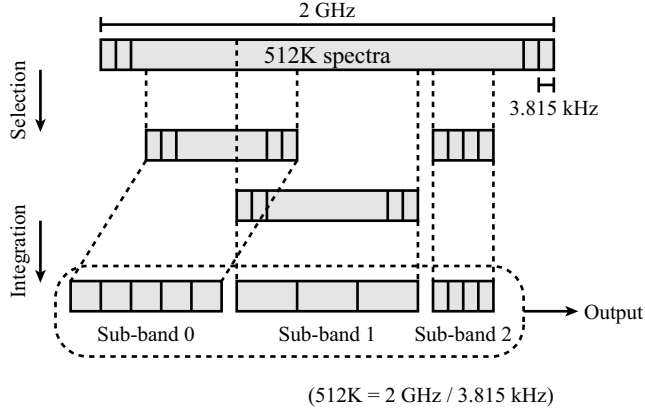}
 \end{center}
 \caption{Up to 32 sub-bands are available to select frequency ranges
 from 512K spectra, whose total bandwidth is 2GHz and frequency
 resolution is 3.815 kHz. Selected spectra are integrated up to the
 requested frequency resolutions and then sent to the output.}
 \label{fig:acafx_sub-band}
\end{figure}

After the correlation multiplication, spectra of requested frequency
ranges corresponding to sub-bands are selected and then integrated up to
the requested frequency resolutions (see Figure
\ref{fig:acafx_sub-band}). The maximum number and bandwidth of sub-bands
are 32 and 2 GHz, respectively. The available spectral integration is
1, 2, 4, 8, 16, 32, 64, 128, 256, 512 and 1024 frequency
channels $\times$ 3.815 kHz. If integrations larger than 1024
frequency channels $\times$ 3.815 kHz are needed, further
integration is performed by post-processing computers.
In the spectral integration, two types of integration methods are
available. One is non-weighted spectral integration, which just adds up
spectral data without weighting in the frequency domain. The other is
frequency profile synthesis, which performs the convolution of weighting
function in the frequency domain. This function is prepared for the
frequency profile compatibility with the 64-Antenna Correlator. The ACA
Correlator and the 64-Antenna Correlator adopt different correlation
methods, which are FX and XF, respectively. These two correlation
methods provide different frequency responses for their correlation
spectra. However, the combined imaging of the ACA and the 12m-Array data
is required in ALMA. Hence, we implement the frequency profile synthesis
function to retain the frequency profile compatibility. The detailed
analysis and implementation are described in \citet{Kamazaki2008a}.


In the temporal integration with the ACA Correlator, integration time of
16 ms and 1 ms are prepared for interferometric observations (auto- and
cross-correlation spectra of all the antennas and baselines) and
single-dish observations (auto-correlation and cross-polarization
spectra of all antennas), respectively.
The spectral and temporal integrations satisfy the ALMA system
requirements and allow the compatibility with the 64-Antenna Correlator.

\section{Hardware implementation}
\label{sec:hw_implementation}

We have implemented the designed functions on the ACA Correlator, whose
specifications are summarized in Table \ref{tbl:acafx_spec}.


\begin{table}
 \caption{Detailed specifications of the ACA Correlator}
 \label{tbl:acafx_spec}
 \begin{center}
  \begin{tabular}{ll}
   \hline \hline
   Input & \\
   \hline
   Number of antennas   & 16 \\
   \hline
   \raisebox{1.5ex}{Number of inputs per antenna}
   & \shortstack[l]{
       \\ 4 baseband pairs $\times$ 2 polarizations \\ = 8 basebands} \\
   \hline
   Processing bandwidth per input & 2 GHz \\
   \hline
   Sampling speed per input & 4 Gsps \\
   \hline
   Number of bits per sample & 3 bits \\
   \hline
   Number of optical input & 12 per antenna, 192 in total \\
   \hline
   Timing signal & 48 ms Timing Event signal \\
   \hline \hline
   \multicolumn{2}{l}{Output per auto-/cross-correlation} \\
   \hline
   \shortstack[l]{
       \\ Maximum number of \\ processed correlation \\ per baseband pair}
   & \raisebox{1.5ex}{\shortstack[l]{
       120 cross-correlations \\ 16 auto-correlations}} \\
   \hline
   \raisebox{1.5ex}{Polarization}
   & \shortstack[l]{
       \\ (XX), (YY), (XX,YY) \\ or (XX,YY,XY,YX)} \\
   \hline
   \shortstack[l]{
   \\ Maximum number of \\ sub-bands}
   & \raisebox{1.5ex}{32 for observations} \\
   \hline
   \shortstack[l]{
   \\ Maximum number of \\ frequency channels}
   & \shortstack[l]{
       \\ 8192 frequency channels \\ (including polarizations)}\\
   \hline \hline
   Function & \\
   \hline
   Correlation method & FX \\
   \hline
   Number of sub-arrays & 2 \\
   \hline
   Maximum delay compensation & 15 km \\
   \hline
   \raisebox{1.5ex}{Sideband rejection}
   & \shortstack[l]{
       \\ 90-degree phase switching \\ or LO offset} \\
   \hline
   Sideband separation & 90-degree phase switching \\
   \hline
   \raisebox{1.5ex}{Highest frequency resolution}
   & \shortstack[l]{
       \\ 3.815 kHz \\ (= 2 GHz / 512K frequency channels)} \\
   \hline
   Spectral integration
   & $2^{n} (n=0,1,2,...,10) \times 3.815 \mbox{kHz}$ \\
   \hline
   Frequency profile synthesis
   & Yes \\
   \hline
   \raisebox{1.5ex}{Temporal integration}
   & \shortstack[l]{
       \\ 1 ms (auto-correlation only)
       \\ 16 ms (auto- and cross-correlation)} \\
   \hline \hline
   Environment conditions & \\
   \hline
   Temperature & 22 - 28 $^{\circ}$C \\
   \hline
   Humidity & 10 - 90 \% (No condensation) \\
   \hline
   Altitude & AOS (5100 m) \\
   \hline \hline
   EMC & CISPR Class B \\
   \hline \hline
  \end{tabular}
 \end{center}
\end{table}

\subsection{Basic concepts of hardware design}

For the hardware design, we adopted five policies: 1)
distributed processing; 2) use of common modules and card structures; 3)
use of optical transmission in long-distance data transmission; 4)
efficient cooling; and 5) distributed power supplies.
Distributed processing allows performance reduction of each processor
sharing the task and enables us to use simple but general-purpose
processing chips such as FPGAs with a low
frequency. This also allows distributed heat sources, which has a merit
in thermal dissipation.
The use of common modules and card structures reduces the number of
case/card types, and realizes simple mounting and accordingly simple
maintenance operations.
Optical transmission is more reliable than electrical in the bulk/long
distance data transmission within the ACA Correlator (e.g. 500 Gbps data
in total are transferred from the F-part to the X-part) and between the
ACA Correlator and post-processing computers.
The ACA Correlator is installed in the Array Operation Site (AOS)
technical building at an altitude of about 5100 m. Cooling function is
designed to maintain sufficient efficiency in an environment where only
half as much air is available as at sea level.
Stable electric power is supplied to each electrical component
by ``DC power generation and voltage conversion'' placed close to the
components.

\begin{figure}
 \begin{center}
  \FigureFile(122mm,85mm){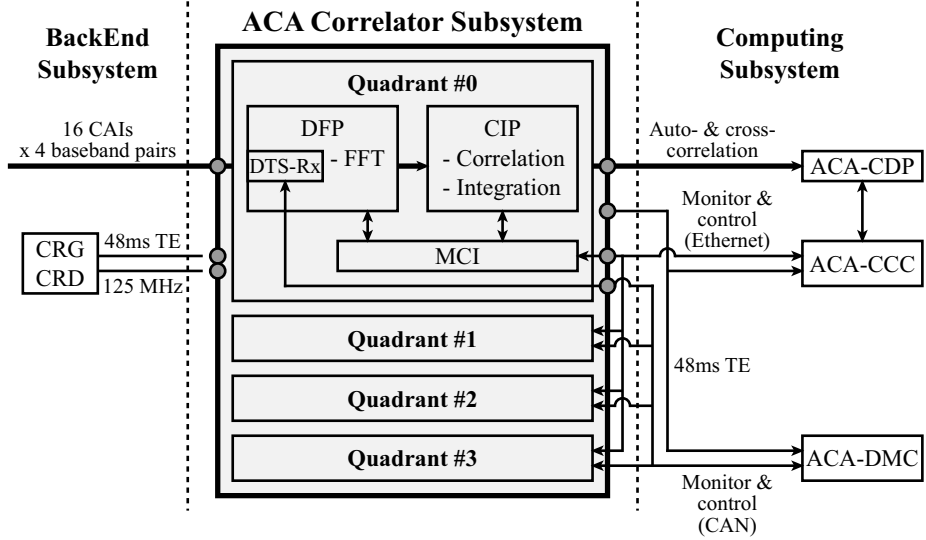}
 \end{center}
 \caption{System block diagram of the ACA Correlator subsystem.
The correlator is controlled by the ACA Correlator Control
 Computer (ACA-CCC) through Ethernet. Optical data receivers on the DFP
 modules are controlled by the ACA Data transmission system receiver
 Monitor and Control computer (ACA-DMC) to maintain the compatibility
 with the 64-Antenna Correlator. Processed data are output from the CIP
 modules to the ACA Correlator Data Processor (ACA-CDP) for their
 post-processing. 48 ms timing event and 125 MHz reference
 signals are supplied from the Central Reference Generator (CRG) and the
 Central Reference Distributor (CRD) in the Backend subsystem,
 respectively.}
 \label{fig:acafx_subsystem}
\end{figure}

Based on these policies, we have designed the ACA Correlator hardware as
follows: The ACA Correlator consists of four quadrants, each of which
processes a baseband pair. The quadrant is mainly divided into three
parts: Data Transmission System Receiver and FFT Processor (DFP)
modules; Correlation and Integration Processor (CIP) modules; and
Monitor \& Control Interface (MCI) module, as shown in Figure
\ref{fig:acafx_subsystem}.
The DFP module, corresponding to the F-part of the FX-method correlator,
receives optical signals transmitted from each antenna, restores
observation data, and performs FFT processing of the restored data.
The CIP module, corresponding to the X-part of the FX
correlator, performs correlation multiplication among voltage spectra
(FFT outputs), and outputs auto- and cross-correlation spectra
at various temporal and spectral resolutions.
The MCI module is responsible for monitoring and control of the DFP and
CIP modules in a quadrant, and has the function to interface
with a control computer as well.

\begin{figure}
 \begin{center}
  \FigureFile(85mm,85mm){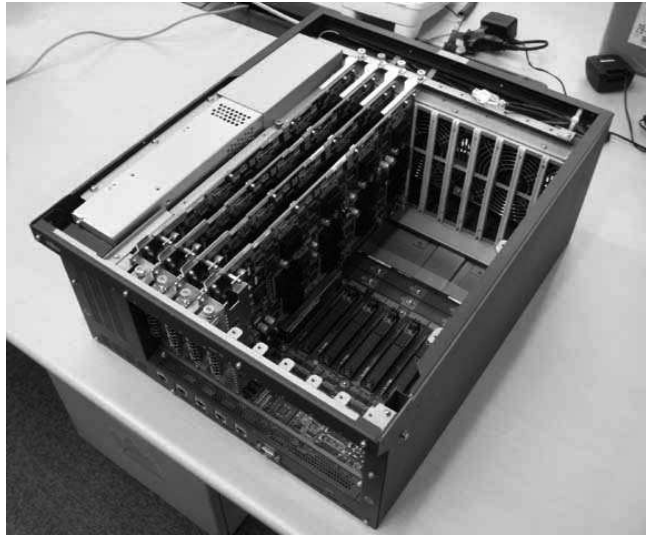}
 \end{center}
 \caption{Inside of the DFP/CIP/MCI module. The DFP/CIP/MCI modules have
 a common structure, which consists of a motherboard, processing cards,
 a power supply unit and a cooling fan unit. Note that all card
 slots are normally filled with processing cards.}
 \label{fig:module}
\end{figure}

\begin{figure}
 \begin{center}
  \FigureFile(85mm,85mm){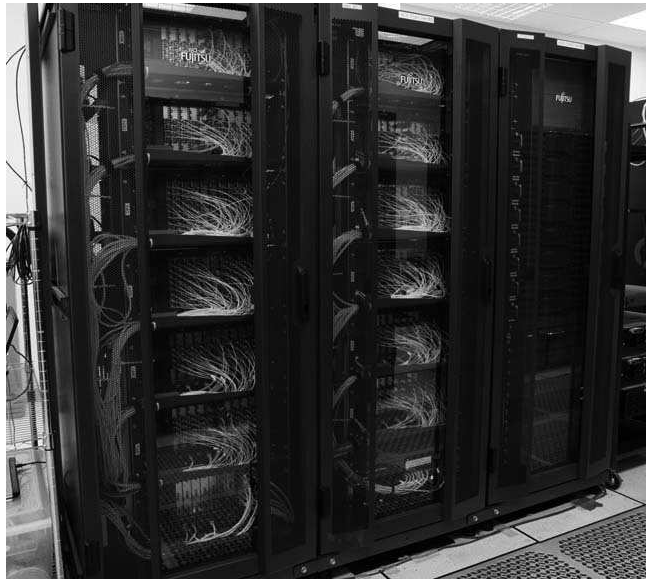}
 \end{center}
 \caption{A quadrant of the ACA Correlator. The left and middle racks
 compose a quadrant, which processes a baseband pair from sixteen
 antennas. The right rack is one of four computing racks, which
 have correlator control, post processing computers, network instruments
 and so on.}
 \label{fig:quadrant}
\end{figure}

These three types of modules adopt a common module/card structure, which
consists of a motherboard, processing cards such as FFT and correlation
multiplication, a power supply unit and a cooling fan unit (see Figure
\ref{fig:module}). The processing cards are aligned in parallel within a
module so that the air entering from the front side of a module goes
straight through a module, and is exhausted from the back side.
Each module is equipped with a switching power supply unit for AC-DC
conversion from AC230 V to DC12 V. The generated DC12 V is converted
to an appropriate voltage (e.g. DC1.2 V for FPGAs) by DC-DC
converters, which are distributed to a motherboard and cards in order to
supply necessary electric power to each electric component at a short
distance. We adopted this method, because it is difficult to
send stable electric power with low voltage and high current
(e.g. DC1.2 V/80 A for a FFT card) over a long distance.
Optical transmission is used to realize reliable transmission of bulk
data (3.125 Gbps per optical fiber) between the modules over a long
distance (up to about 3 m). For the communications between the MCI
module and the other modules, 100 Base-T Ethernet is used through a
network switch.
For remote operations, the electric power of the correlator can be
controlled using network outlets which are controllable from computers
through Ethernet.
48-ms timing signal and 125 MHz reference signal from the Backend
subsystem are received at the Reference Signal Distributor (RSD) panel,
and are distributed to each quadrant through the panel.
As a result, a quadrant of the ACA Correlator is composed of eight DFP
modules, four CIP modules, one MCI module, one RSD panel (installed only
in the first quadrant), one Ethernet controlled outlet and one network
switch. Figure \ref{fig:quadrant} shows a quadrant of the ACA Correlator
installed at the AOS.


\subsection{DFP module}

\begin{figure}
 \begin{center}
  \FigureFile(85mm,85mm){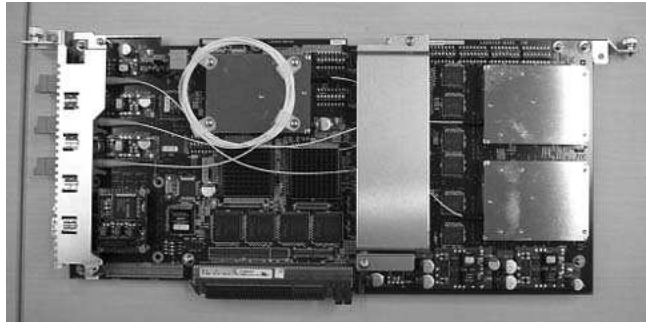}
 \end{center}
 \caption{DTS-Rx card. Three metal boxes connected with optical fibers
 are 1/2 transponders. Three FPGAs, which are under a metal plate to fix
 the optical fibers, control the transponders. Two FPGAs with heat sinks
 are DFR parts.}
 \label{fig:dts-rx_card}
\end{figure}

The DFP module has two types of processing cards, which are Data
Transmission System Receiver (DTS-Rx) cards (Figure
\ref{fig:dts-rx_card}) for optical data reception and FFT cards (Figure
\ref{fig:fft_card}) for 1M-point FFT operation. One DFP module has two
DTS-Rx cards and eight FFT cards.

\subsubsection{DTS-Rx card}

The DTS-Rx card is composed of two parts. One is the DTS (Data
Transmission System) fiber optical receiver (FOR), and the other is the
DTS deformatter (DFR).
FOR receives 3-bit 4 Gsps optical signals from the antennas, which are
separately transmitted every bit, and converts them into electric
signals. FOR on the DTS-Rx card processes a baseband pair signal of an
antenna. For the reception, a PIN photo diode is used by
taking the necessary input power level into
account. This level is between 0 dBm and -16 dBm.
DFR decodes embedded frame information from the electric signals
according to the procedures described in \citet{Freund2002}. Then, DFR
applies coarse delay compensation (see Section
\ref{subsec:coarse_delay}) to the retrieved data per bit and joins three
1-bit data together into one 3-bit data stream. Finally, a
baseband pair is separated into two basebands by polarization, and, if
necessary, data acquisition for HFS detection (see Section
\ref{subsec:spurious_signal_cancellation}) are performed and FFT
segments of $2^{20}$ data are constructed. The FFT segments are combined
in a data transfer unit, which is 4 FFT segments $\times$ 2 baseband
pairs $\times$ 2 antennas, and are transmitted to the FFT cards.


\subsubsection{FFT card}

\begin{figure}
 \begin{center}
  \FigureFile(85mm,85mm){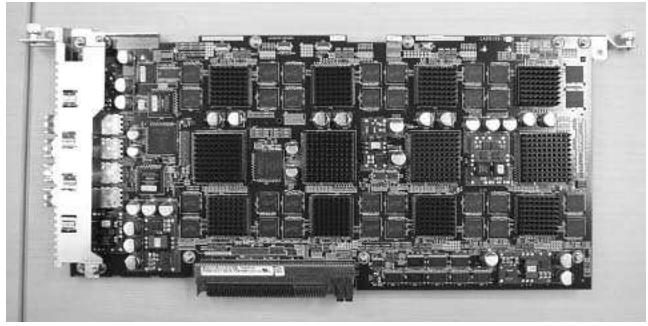}
 \end{center}
 \caption{FFT card. One FFT card is equipped with four FPGAs for control
 (the largest FPGAs with heat sinks) and eight FPGAs for FFT (the second
 largest FPGAs with heat sinks). One card processes 64 FFT segments in
 every 32 ms.}
 \label{fig:fft_card}
\end{figure}

The FFT card is equipped with two types of FPGA: FPGA for FFT and FPGA
for control (see Figure \ref{fig:fft_card}).
The FFT FPGA (Xilinx Virtex-4 LX60) performs 1M-point FFT and related
processing, HFS cancellation and LFS detection \& cancellation before
FFT (see Section \ref{subsec:spurious_signal_cancellation}), residual
delay compensation (see Section \ref{subsec:residual_delay}),
re-quantization (see Section \ref{subsec:re-quantization}) and LO offset
demodulation (see Section \ref{subsec:lo_offset}).
On the other hand, the control FPGAs (Xilinx Virtex-4 LX80) allocate the
segmented data transferred from the DTS-Rx cards to the FFT FPGAs, and
transmit FFT outputs collected from the FFT FPGAs to the CIP
modules. For these communications, the control FPGA requires more
input/output pins and adopts larger FPGA than the FFT FPGA. The data
transfer is operated in units of 2 antennas $\times$ 2 polarizations
$\times$ 512K 4-bit complex $\times$ 4 FFT segments.

\begin{figure}
 \begin{center}
  \FigureFile(93mm,85mm){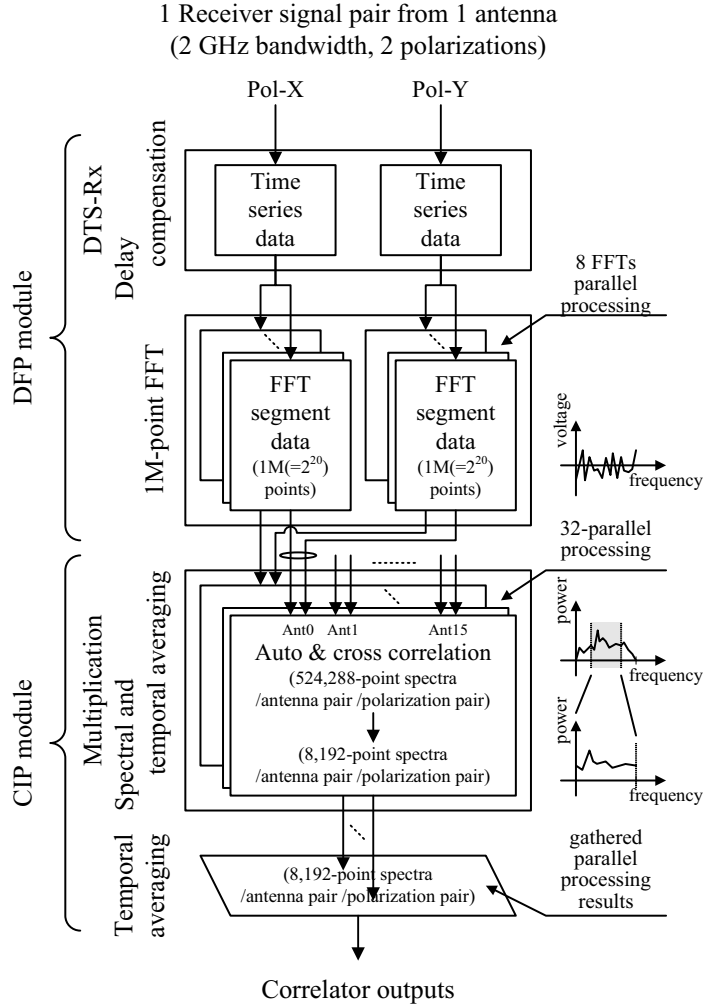}
 \end{center}
 \caption{The detailed processing of the ACA Correlator. Time-series
 data of each antenna are divided in the time domain and processed in
 parallel.}
 \label{fig:acafx_proc}
\end{figure}

One FFT card has eight FFT FPGAs. A pair of the FPGAs processes 1M-point
FFT operation of 16 FFT segments, corresponding to a unit of data
transfer from the DTS-Rx card, every 32 ms. Hence, one FFT card
processes 64 FFT segments, and 512 FFT operations in total are performed
by eight FFT cards of a DFP module every 32 ms. The processing is
performed in parallel by eight pairs of the FPGAs within a DFP module as
shown in Figure \ref{fig:acafx_proc}. Each FPGA pair is responsible for
FFT segments with different time slots.
This parallel processing in the time domain has advantages in design and
manufacturing, because common data processing can be applied to all
parallel processors and uniform design and manufacturing procedure can
be adopted. If different processes are in parallel (e.g. butterfly
operations and twiddle factor multiplication in FFT), we have to design
and manufacture them individually. The parallel processing has also an
advantage in thermal dissipation. Since it allows one FFT card to take
longer processing time than single processing, we can reduce power
consumption of each FPGA by decreasing the clock frequency to
drive them. The total power consumption of the whole system does not
change much because of the increase in the number of FPGAs. However, we
can avoid local heat concentration. The distributed
heat-sources with lower power consumption lead to the reduction of
cooling fans; small fans attached to each FPGA can be replaced with a
few larger fans that are common to FPGAs and consume less electricity in
total. Furthermore, we can reduce failures associated with fan
failures.
Another advantage is the flexibility of spectral processing. One FFT
card can easily access all 512K-point voltage spectra, which
are halves of 1M-point FFT outputs, because spectral data are
distributed to FFT cards in the time domain instead of frequency
domain.

After the FFT operations, voltage spectra are compensated for residual
delay, scaled to the appropriate input level of the 4-bit
re-quantizer and re-quantized to 4-bit (this scaling is
compensated after the correlation multiplication in the CIP modules).
Finally, the 4-bit re-quantized voltage spectra are shifted in the
frequency channel domain for LO offset demodulation if required, and
then, transferred to the CIP modules.


\subsection{CIP module}

\begin{figure}
 \begin{center}
  \FigureFile(85mm,85mm){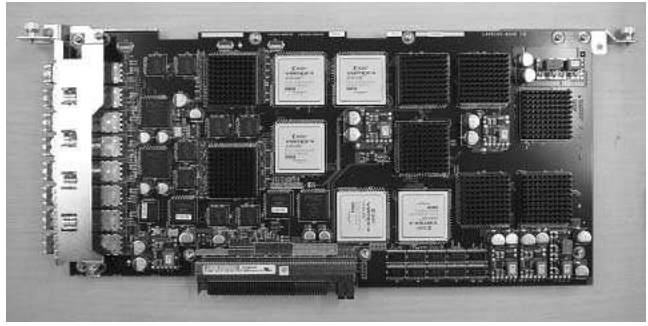}
 \end{center}
 \caption{CIP card. Each card calculates power spectra in all
 combinations of the voltage spectra. The power spectra are
 integrated to 1 ms in the time domain and to 
 half the output frequency spacing in the frequency domain.}
 \label{fig:cip_card}
\end{figure}

\begin{figure}
 \begin{center}
  \FigureFile(85mm,85mm){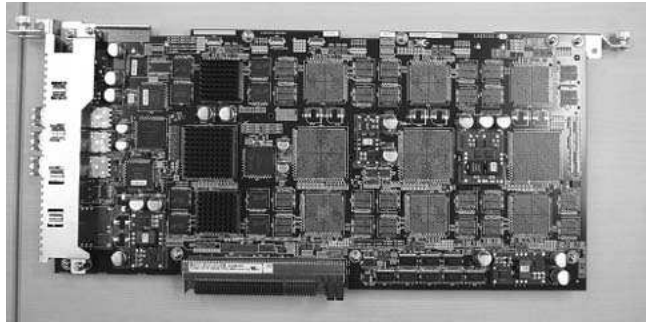}
 \end{center}
 \caption{MTI card. Further temporal and spectral integration are
 performed in this card. The MTI card uses the same circuit boards as
the  FFT card in order to minimize the number of different
 circuit patterns. A smaller number of FPGAs is installed on the MTI
 card, because the processing amount is smaller compared to the FFT card.}
 \label{fig:mti_card}
\end{figure}

In the CIP module, there are two types of processing cards: CIP cards
(Figure \ref{fig:cip_card}) for correlation multiplication; and MTI
cards (Figure \ref{fig:mti_card}) for data output. One CIP module has
eight CIP cards and two MTI cards.

\subsubsection{CIP card}

One CIP module has eight CIP cards, each of which is connected with a
FFT card in each of eight DFP modules by one-to-one optical
connections. This means that each CIP card receives all voltage spectra
of dual polarizations of all antennas with different time slots
and processes them in parallel with 32 CIP cards (= 4 CIP modules
$\times$ 8 CIP cards) as shown in Figure \ref{fig:acafx_proc}. Each CIP
card can obtain power spectra in all combinations of the voltage spectra
by correlation multiplication and easily access all spectra of
full bandwidth (2 GHz or 524,288 frequency channels). Hence, it is
available to realize flexible frequency processing such as multiple
sub-bands and multiple resolutions (e.g. coarse spectra of 2 GHz full
bandwidth and fine spectra of maser emission in close-up). Only the
power spectra of polarization pairs and frequency ranges required for
users are integrated to 1 ms (corresponding to four FFT segments) in the
time domain and to half the output frequency spacing in the
frequency domain.

The parallel processing results are transferred to MTI cards through
electric serial bus dedicated to bulk data transfer. Its maximum data
rate is still 16K-point 20-bit power spectra per antenna/baseline after
the temporal and spectral integration, corresponding to about 2.8 Gbps,
although available transfer rate of the serial bus is 2.5 Gbps for
actual data. In addition, 20-bit or more accuracy is needed for the
transfer in order to achieve calculation bias less than -60 dB relative
to $T_{SYS}$, which is required for spectral dynamic range of
10000:1. Thus, ``Differential transfer method'', or ``$\Delta\Sigma$
transfer method'', is adopted to realize both data-rate
reduction and effective 20-bit data transfer.

In the differential transfer, only the difference between adjacent data
in the frequency domain are transmitted and received except for the
first data of each sub-band. Here, the difference is taken between
``present modulator input'' and ``previous demodulator output'' instead
of ``previous modulator input'' in order to avoid permanent demodulation
output error after a difference overflow due to the limited word length
of the transmitted data. Thus, when the spectra are sufficiently smooth
by scaling at the re-quantization, the bit rate of the transfer
decreases dramatically.


\subsubsection{MTI card}

Within a CIP module, correlation spectra are integrated to 1 ms
in each of the eight CIP cards and are transferred to two MTI
cards. The MTI card performs further temporal integration to 4 ms,
rescaling to compensate for scaling at the 4-bit
re-quantization in the DFP modules and frequency profile
synthesis/remaining non-weighted spectral integration. Further temporal
integration to 16 ms is performed while the spectra are being passed to
one another among four MTI cards of two CIP modules. Final correlation
spectra are distributed to eight MTI cards on four CIP modules
by a similar bucket brigade method and are output to
ACA-CDPs. In this transfer between MTI cards, normal data transfer is
performed because data rate is decreased by further temporal and
spectral integration.


\subsection{MCI module}

The MCI module is an interface with the ACA Correlator control
computer. The ACA-CCC controls the ACA Correlator with User Datagram
Protocol (UDP) through Ethernet. We adopt UDP because time-sensitive
control is required for delay compensation, processing start and stop,
and other operations, although its reliability is lower than
Transmission Control Protocol (TCP) in terms of error correction.
The MCI module contains cards for power control and reference signal
distribution. The card controls power-on and power-off of other
modules. When the MCI module is turned on, it starts to power on other
modules sequentially in order to avoid large rush current. When the MCI
module is turned off, it immediately turns off other modules. Since electric
power for a MCI module is supplied through a network controllable
outlet, an operator can remotely power-on and power-off the ACA Correlator by
controlling the outlet. The card also distributes 48 ms timing clock
to other modules. The clock is used as base clock in the ALMA system and
is provided from a central reference distributor.

\section{Verification of the ACA Correlator}
\label{sec:verification}

We verified the design and actual implementation of the ACA Correlator
by conducting functionality, stability and performance
tests. In the tests, a known digital input signal was used to
the ACA Correlator with typical configuration parameters, and its
outputs were checked to confirm that they were consistent with 64-bit
floating-point arithmetic results for the design verification, and/or
bit-accurately coincident with the ACA Correlator simulator results for
the implementation verification. In the performance test, 1-bit
digitized thermal noise was also input in order to evaluate upper limits
of calculation bias and sensitivity loss that occurred within the ACA
Correlator.

The 64-bit floating-point arithmetic was performed using Interactive
Data Language (IDL; http://www.ittvis.com/ProductServices/IDL.aspx) on
Linux PCs. We used the FFT function implemented on IDL for 1M-point FFT
calculations. Other functions were realized by 64-bit floating-point
addition, subtraction, multiplication, division and their
combinations. Since all correlation calculations are processed
by using a simple FX operation with 64-bit floating-point arithmetic, an
ideal correlation result can be expected at the output. Thus, by
checking that the correlator results are consistent with the 64-bit
floating-point arithmetic results, we can verify that our design is
correct.

However, we cannot check that implemented processing is identical with
what we designed by using the 64-bit floating-point arithmetic, because
detailed calculation methods (e.g. number rounding) are different
between the arithmetic and the correlator hardware. For the purpose of
the implementation verification, we developed an ACA Correlator
simulator (hereafter C++ simulator), which is a bit-accurate software
simulator written in C++ by using FFT library provided by Xilinx
(http://www.xilinx.com/products/ipcenter/FFT.htm). Since this simulator
is equipped with most of the functions of the actual ACA Correlator such
as FFT, correlation multiplication, re-quantization, and integration in
the time and frequency domain and accepts the same digital data as the
ACA Correlator, it can calculate exact values equivalent to the
ACA Correlator. Thus, by checking if the correlator results
are bit-accurately coincident with simulator results, we can verify that
the implemented processing is identical with what we designed.

\subsection{Functional tests}
\label{subsec:function_tests}

At first, we conducted functional tests to check the basic
processing functions of the ACA Correlator: FFT; re-quantization;
correlation multiplication; and spectral and temporal integration.

\begin{figure}
 \begin{center}
  \FigureFile(85mm,85mm){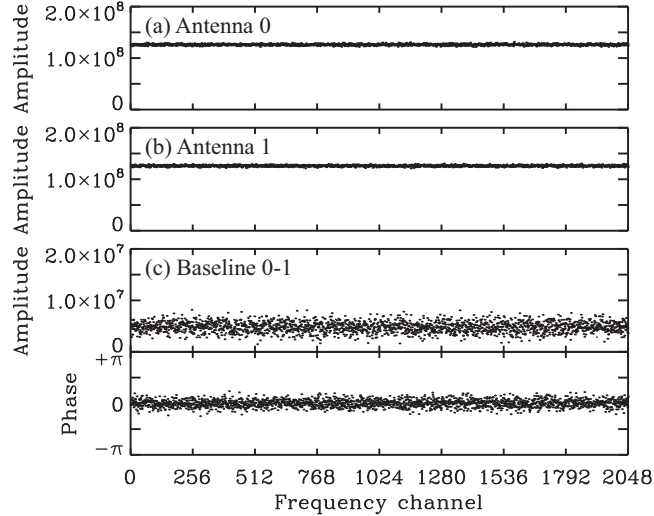}
 \end{center}
 \caption{
 (a) Auto-correlation spectra of Antenna 0.
 (b) Auto-correlation spectra of Antenna 1.
 (c) Cross-correlation spectra between Antenna 0 and Antenna 1.
 The bandwidth and channel spacing are 2 GHz and 244 kHz, respectively.
 The integration time is 16 ms.}
 \label{fig:sft03x}
\end{figure}

Figure \ref{fig:sft03x} shows an example of the functional
tests, where computationally generated 3-bit correlated
Gaussian white noises were used as input to the ACA
Correlator. The input data were common to almost all the
antennas except for Antenna 1. The data correlated with the common input
was used for Antenna 1. The amplitude plots of the auto- and
cross-correlation spectra show flat spectra. The phases of the
cross-correlation spectra distribute around 0 radian. These are as
expected from the input data.
Correlation coefficient $\rho$ of Figure \ref{fig:sft03x}c is estimated
to be 0.0374 by the following equation:
\begin{eqnarray}
 \rho & = & \frac
  {\left |\sum_{f=0}^{(2048-1)} C_{01}(f) \right|}
  {\sqrt{\sum_{f=0}^{(2048-1)}A_{00}(f) \sum_{f=0}^{(2048-1)}A_{11}(f)}} \\
 C_{01} & : & \mbox{cross-correlation spectra of Baseline 0-1} \nonumber \\
 A_{00} & : & \mbox{auto-correlation spectra of Antenna 0} \nonumber \\
 A_{11} & : & \mbox{auto-correlation spectra of Antenna 1} \nonumber \\
 \rho   & : & \mbox{correlation coefficient} \nonumber \\
 f      & : & \mbox{frequency channels} \nonumber
\end{eqnarray}
This is consistent with 0.0374, which is calculated by 64-bit
floating-point arithmetic from the same input data. These facts confirm
that our algorithm design of the basic processing functions of
the ACA Correlator are correct.
In addition, we also confirmed that all correlation spectra are
bit-accurately coincident with the C++ simulator results. Thus, it is
verified that the basic processing is implemented as designed.


\begin{figure}
 \begin{center}
  \FigureFile(85mm,85mm){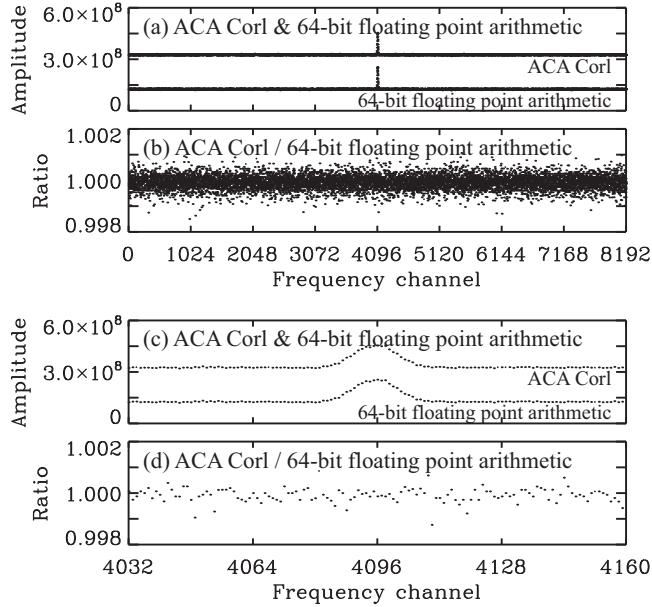}
 \end{center}
 \caption{
 (a) ACA Correlator and 64-bit floating point arithmetic results.
 (b) The ratio of the ACA Correlator results to 64-bit floating point
 arithmetic results.
 (c) Magnified plot of the band center of the plot (a).
 (d) Magnified plot of the band center of the plot (b).
 The bandwidth and channel spacing are 2 GHz and 244 kHz, respectively.
 The integration time is 16 ms.}
 \label{fig:sft04a}
\end{figure}

Figure \ref{fig:sft04a}a shows an example of a comparison of
auto-correlation spectra between 64-bit floating-point arithmetic and
ACA Correlator results. Input signal is a mixture of Gaussian
like profile signal and Gaussian white noise, which are computationally
generated. Figure \ref{fig:sft04a}c is a plot which zooms in on the
band center of Figure \ref{fig:sft04a}a. These amplitude plots indicate
that they are well consistent with each other, as shown in the ratio
plots (Figure \ref{fig:sft04a}b and \ref{fig:sft04a}d) between them. The
mean, standard deviation, maximum and minimum of the ratio is 0.999921,
0.000271, 1.001065 and 0.998501, respectively. Thus, they are coincident
with each other within the accuracy of the measurements, and our basic
design of the processing functions has been confirmed to be correct.
Note that the bit-accurate coincidence is also confirmed in this test.


\begin{figure}
 \begin{center}
  \FigureFile(85mm,85mm){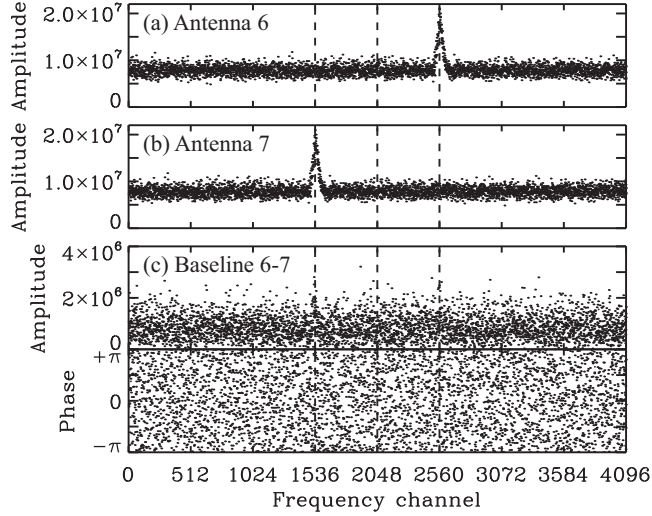}
 \end{center}
 \caption{
 (a) Auto-correlation spectrum of Antenna 6.
 (b) Auto-correlation spectrum of Antenna 7.
 (c) Cross-correlation spectrum between Antenna 6 and Antenna 7.
 The bandwidth and channel spacing are 31.25 MHz and 3.815 kHz,
 respectively.} The integration time is 16 ms.
 \label{fig:sft56a}
\end{figure}

Figure \ref{fig:sft56a} shows an example of other functional
tests. This test was performed to check the LO offset function (see also
Section \ref{subsec:lo_offset}). Input signal was computationally
generated by mixing Gaussian-like profile signal, which was located in
the center of 2 GHz bandwidth, with Gaussian white noise. Thus, the
Gaussian-like profile is expected to be seen at the frequency channels
offset from the band center by LO offset parameters. Figure
\ref{fig:sft56a}a and Figure \ref{fig:sft56a}b show the signal shift to
2560 and 1536 frequency channels from 2048 frequency channel according
to LO offset parameters, which are +512 and -512 of Antenna 6 and
Antenna 7, respectively. The frequency channel shifts are also confirmed
by binary comparison between the data with LO offset and without LO
offset. As a result of the opposite frequency shifts, no Gaussian-like
profile signal is seen in cross-correlation between Antenna 6 and
Antenna 7 as shown in Figure \ref{fig:sft56a}c.


\subsection{Stability tests}


The data reception and processing stability of the ACA Correlator were
verified by running the correlator for eight hours. In the stability
tests, known test data was repeatedly input to the correlator through
optical fibers, and output correlation products were checked to see that
they were bit-accurately coincident with C++ simulator results in
real-time.
Table \ref{tbl:sft8687} shows the results of the two tests, which were
conducted in the full-array and sub-array modes. This can be interpreted
that all input data were successfully processed and all output
correlation products were correct for eight hours.


\begin{table}
 \caption{Stability test results}
 \label{tbl:sft8687}
 \begin{center}
  \begin{tabular}{ccc}
   \hline
   Mode & Full-array & Sub-array \\
   \hline
   \hline
   \raisebox{0.5ex}{\shortstack[c]{Number of \\ received data \\ at
   optical receivers}}
   & \shortstack[c]{
       \\
       4 Gbps $\times$ 3-bit \\
       $\times$ 2 polarizations \\
       $\times$ 16 antennas $\times$ 8 hours}
   & \shortstack[c]{
       \\
       4 Gbps $\times$ 3-bit \\
       $\times$ 2 polarizations \\
       $\times$ 16 antennas $\times$ 8 hours} \\
   \hline
   \shortstack[c]{ \\ Bit error rate \\ at data reception}
   & \raisebox{1.0ex}{$<9.0 \times 10^{-17}$}
   & \raisebox{1.0ex}{$<9.0 \times 10^{-17}$} \\
   \hline
   \hline
   \raisebox{2.5ex}{\shortstack[c]{Number of output \\ correlation products}}
   & \shortstack[c]{
       \\
       8192 frequency channels \\
       $\times$ (16 antennas \\
       + 120 baselines) \\
       $\times$ 1800000 blocks}
   & \shortstack[c]{
       \\
       8192 frequency channels \\
       $\times$ (16 antennas \\
       + 66 baselines) \\
       $\times$ 1800000 blocks} \\
   \hline
   \shortstack[c]{ \\ Number of \\ incorrect data}
   & \raisebox{1.0ex}{0}
   & \raisebox{1.0ex}{0} \\
   \hline
  \end{tabular}
 \end{center}
\end{table}

\subsection{Performance tests}

As described in Section \ref{sec:specs_reqs}, the ACA Correlator is
required to achieve a spectral dynamic range of 10000:1 and
sensitivity loss less than 2-bit quantization, which corresponds to 12
\%. Since the sensitivity loss includes 3-bit digitization, whose loss
is 3.7 \%, the available error budget for the
correlator is 8.3 \%. Hence, we designed the calculation bias and
sensitivity loss added within the ACA Correlator to be less than
1/10$^{6}$ of the system temperature $T_{SYS}$ and 8.3 \%,
respectively. However, their verification by actual tests is difficult,
especially that of spectral dynamic range, within a finite time.
The number of combinations of available correlator configurations is
almost infinite, because there are many parameters in sub-arrays,
sub-bands, polarizations, temporal integration time, delay, phase
switching, LO offset and so on. In addition, the integration time
necessary for the verification of the spectral dynamic range is about
3000 days with the following equation assuming the highest frequency
resolution of 3.815 kHz and the strong line emission of 1/100 of
$T_{SYS}$. Thus, we verified the requirement by analytical and
simulation study, which will be reported elsewhere.
\begin{displaymath}
 \frac{T_{SYS}}{\sqrt{\mbox{3.815 kHz} \times \mbox{[integration time (sec)]}}}
 = \frac{1}{10000} \frac{T_{SYS}}{100}
\end{displaymath}
In actual tests, we checked the following two items within the limited
correlator configurations and available integration time. The first is
that calculation bias is smaller than the accuracy of the measurements
and is not significantly detected. The second is that additional
sensitivity loss caused by the correlator is consistent with that by
4-bit re-quantization within the accuracy of the measurement
and is smaller than the allowable loss of 8.3 \%.

\subsubsection{Calculation bias}

\begin{figure}
 \begin{center}
  \FigureFile(85mm,85mm){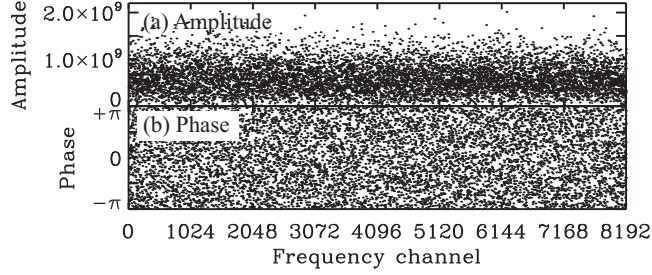}
 \end{center}
 \caption{
 Amplitude (a) and phase (b) of cross-correlation spectra of two 1-bit
 digitized incoherent thermal noises over eight-hour
 integration. The bandwidth and frequency resolution are 31.25
 MHz and 3.815 kHz, respectively.}
 \label{fig:spt01b_ch}
\end{figure}

\begin{figure}
 \begin{center}
  \FigureFile(85mm,85mm){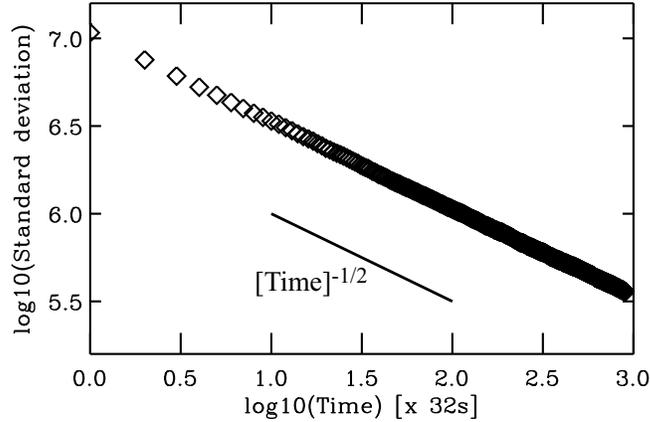}
 \end{center}
 \caption{
 Rms of amplitude of cross-correlation spectra over 8192
 frequency channels as a function of the total integration time. The
 overplotted line shows the inclination of $1/\mbox{[integration
 time]}^{-1/2}$.}
 \label{fig:spt01b_tm}
\end{figure}

We verified the calculation bias by using two 1-bit digitized
incoherent thermal noises as input for the ACA Correlator and
accordingly integrating cross-correlation products over eight hours. If
the correlator internally generates any artificial biases larger than
$T_{SYS} / \sqrt{3.815 kHz \times \mbox{8 hours}}
\sim T_{SYS} / 10^{4}$,
it is expected that such biases appear in amplitude and phase plot and,
that noise level does not fall below the level determined by
$\sim T_{SYS} / 10^{4}$.
Figure \ref{fig:spt01b_ch} shows the amplitude and phase plots of
cross-correlation spectra integrated over eight hours. These plots
clearly show that there is neither harmful sporadic event caused by
digital processing nor harmful spurious artificial signal. Note
that the two input signals had weak correlation with each other. This
was caused by signal leakage within the analog noises generated. The LO
offset was used to cancel this correlated component.
Figure \ref{fig:spt01b_tm} shows the root-mean-square (rms) of amplitude
of 8192 cross-correlation spectra as a function of the total integration
time. It is clear that the noise level decreases in proportion to
$1/\mbox{[integration time]}^{-1/2}$.
These results indicate that the calculation bias added in the ACA
Correlator is smaller than $T_{SYS} / 10^{4}$, the level obtained in
eight hours integration.


\subsubsection{Sensitivity loss}

Sensitivity loss within the ACA Correlator has been evaluated by
using computationally generated Gaussian white noises
at the input, whose effective duration is 512 ms and their correlation
coefficient is 0.104. Table \ref{tbl:spt02b} shows averages and standard
deviations of 2048-channel auto-/cross-correlation spectra, whose
frequency resolution is 3.815 kHz, at the total integration time of 512
ms. Thus, sensitivity loss is estimated to be 0.9 \% by the calculation
of
$(399979 / 1755886) / (404727 / 1792962) - 1.0 = 0.009 \pm 0.001$.
This value is comparable with the sensitivity loss by 4-bit
re-quantization, which is estimated to be 1.1 \% by a Monte Carlo
simulation.
This evaluation result demonstrates that sensitivity loss is dominantly
determined by 4-bit re-quantization as expected and the loss is much
smaller compared to the allowable loss of 8.3 \%.


\begin{table}
 \caption{Amplitude of auto- and cross-correlation spectra}
 \label{tbl:spt02b}
 \begin{center}
  \begin{tabular}{ccccc}
   \hline
   & \multicolumn{2}{c}{64-bit floating-point arithmetic}
   & \multicolumn{2}{c}{ACA Correlator} \\
   \hline
   \hline
   & $\mu$ & $\sigma$ & $\mu$ & $\sigma$ \\
   \hline
   Auto-correlation 0
   & 18113972 & 396372 & 17947951 & 371919 \\
   \hline
   Auto-correlation 1
   & 18120062 & 404172 & 17949807 & 376595 \\
   \hline
   Cross-correlation 0-1
   & 1792962 & 404727 & 1755886 & 399979 \\
   \hline
   \multicolumn{5}{l}{$\mu$ : Average of real/complex 2048 spectral data.} \\
   \multicolumn{5}{l}{$\sigma$ : Standard deviation of 2048 spectral data.} \\
   \multicolumn{5}{l}{The values are in arbitrary unit.} \\
   \multicolumn{5}{l}{Integration time is 512 ms.} \\
  \end{tabular}
 \end{center}
\end{table}

\section{Summary}

We have developed the ACA Correlator, which is a digital
spectro-correlator system with an FX architecture. The ACA
Correlator can process all correlation products including
cross-polarization among up to sixteen antennas. It also realizes
flexible spectral processing such as multiple frequency resolutions and
multiple frequency ranges by adopting the FX method with parallel
processing in the time domain. Furthermore, processing compatibility
with the 64-Antenna Correlator is ensured in data reception, sideband
processing such as phase switching and LO offset, spectral and temporal
integration, and frequency response weighting.

We have verified the design and implementation of the developed
correlator by using known digital data as input signal and
comparing its outputs with outputs of 64-bit floating-point arithmetic
and the C++ simulator. The verification results have confirmed that the
hardware outputs are consistent with the 64-bit floating-point
arithmetic outputs and coincident with outputs from the C++ simulator.

In the verification of the correlator performance, it is
confirmed that there is no significant artificial bias in the spectra
integrated over eight hours and that rms and noise level decreases in
proportion to $1/\mbox{[integration time]}^{-1/2}$. The sensitivity loss
is also confirmed to be much lower than the allowable loss of the ACA
Correlator.

\bigskip

We would like to acknowledge the ALMA project members for supporting the
development and installation of the ACA Correlator. We are especially
grateful to Manabu Watanabe, Munetake Momose and Koh-ichiro Morita for
their technical discussion and constructive comments. Finally, we would
like to express our gratitude to Andre Gunst who provided constructive
comments and suggestions for this paper.


\bigskip



\begin{thebibliography}{}
 \bibitem[Beasley(2006)]{Beasley2006}
 Beasley, A.J., Murowinski, R., and Tarenghi, M.,
 2006, Proc. SPIE, 6267, 626702
 \bibitem[Bunton(2003)]{Bunton2003}
 Bunton, J.,
 2003, ALMA Memo 447
 \bibitem[Chikada et al.(1987)]{Chikada1987}
 Chikada, Y., Ishiguro, M., Hirabayashi, H., Morimoto, M., and
 Morita, K.,
 1987, Proc. IEEE, 75, 1203
 \bibitem[Emerson(2006)]{Emerson2006}
 Emerson, D.T., 2006, ALMA Memo 565
 \bibitem[Escoffier et al.(2007)]{Escoffier2007}
 Escoffier, R.P., Comoretto, G., Wevver, Baudry, A., Broadwell, C.M.,
 Greenberg, J.H., Treacy, R.R., Cais, P., Quertier, B., Camino, P.,
 Bos, A., and Gunst, A.W.
 2007, \aap, 462, 801
 \bibitem[Freund(2002)]{Freund2002}
 Freund R.W., 2002, ALMA Memo 420
 \bibitem[Iguchi and Kawaguchi(2002)]{Iguchi2002}
 Iguchi, S., and Kawaguchi, N., 
 2002, IEICE Trans. Commun., 85, 1806
 \bibitem[Iguchi et al.(2005)]{Iguchi2005}
 Iguchi, S.,
 2005, \pasj, 57, 643
 \bibitem[Iguchi et al.(2008)]{Iguchi2008}
 Iguchi, S., and Okuda, T.,
 2008, \pasj, 60, 857
 \bibitem[Iguchi et al.(2009)]{Iguchi2009}
 Iguchi, S., et al.,
 2009, \pasj, 61, 1
 \bibitem[Kamazaki et al.(2008a)]{Kamazaki2008a}
 Kamazaki, T., Okumura, S.K., Chikada, Y., 2008, ALMA Memo 580
 \bibitem[Kamazaki et al.(2008b)]{Kamazaki2008b}
 Kamazaki, T., Okumura, S.K., Chikada, Y., 2008, ALMA Memo 581
 \bibitem[Okumura et al.(2002)]{Okumura2002}
 Okumura, S.K., Iguchi, S., Chikada, Y., and Okiura, M.,
 2002, PASJ, 54, L59
 \bibitem[Thompson, Moran, and Swenson(2001)]{Thompson2001}
 Thompson, A.R., Moran, J.M., Swenson, G.W. Jr, 2999, Interferometry and
 Synthesis in Radio Astronomy (A Wiley-Interscience
 Publication, New York)
 \bibitem[Weinreb(1963)]{Weinreb1963}
 Weinreb, S.,
 1963, A Digital Spectral Analysis Technique and Its Application to
 Radio Astronomy (RLE Technical Report 412)
\end{thebibliography}
\end{document}